\documentclass[12pt,preprint,iop]{emulateapj}

\usepackage{epsfig, natbib} 
\usepackage{mathrsfs,amssymb}
\usepackage{graphicx,latexsym}

\newcommand{\ang}{{\rm \AA}}

\newcommand{\etal}{{\rm et~al.\/}}
\newcommand{\fesc}{\mbox{$f_{\rm esc}$}}
\newcommand{\fwim}{\mbox{$f_{\rm WIM}$}}

\newcommand{\Hline}[1]{\mbox{H{\footnotesize {#1}}}}
\newcommand{\Halpha}{\Hline{\mbox{$\alpha$}}}

\newcommand{\HI}{{\sc Hi}}
\newcommand{\HII}{{\sc Hii}}

\newcommand{\HIPASS}{{\sc HiPASS}}

\newcommand{\mum}{\mbox{$\mu \rm m$}}
\newcommand{\MHI}{\mbox{${\cal M}_{\rm HI}$}}

\newcommand{\Msun}{\mbox{${\cal M}_\odot$}}

\newcommand{\NII}{{\sc Nii}}

\newcommand{\SFRcrit}{\mbox{${\rm SFR}_{crit}$}}

\slugcomment{Submitted to the Astrophysical Journal, 2010 March 24, Accepted October 11}

\shorttitle{Fate of Ionizing Radiation in Starbursts}
\shortauthors{Hanish \etal}

\begin{document}

\title{A Multiwavelength Study on the Fate of Ionizing Radiation in Local Starbursts}

\author{D.J.\ Hanish\altaffilmark{1,2},
        M.S.\ Oey\altaffilmark{1},
        J.R.\ Rigby\altaffilmark{3},
        D.F.\ de\ Mello\altaffilmark{4},
        J.C.\ Lee\altaffilmark{3}}

\altaffiltext{1}{Department of Astronomy, University of Michigan, 500 Church Street, 830 Dennison, Ann Arbor, MI 48109-1042, $hanish@ipac.caltech.edu$}
\altaffiltext{2}{Spitzer Science Center, California Institute of Technology, MC 314-6, Pasadena, CA 91125}
\altaffiltext{3}{Carnegie Observatories, 813 Santa Barbara St., Pasadena, CA 91101}
\altaffiltext{4}{The Catholic University of America, Washington, DC 20064}

\begin{abstract}
The fate of ionizing radiation is vital for understanding cosmic ionization, energy budgets in the interstellar and intergalactic medium, and star formation rate indicators.  The low observed escape fractions of ionizing radiation have not been adequately explained, and there is evidence that some starbursts have high escape fractions.  We examine the spectral energy distributions of a sample of local star-forming galaxies, containing thirteen local starburst galaxies and ten of their ordinary star-forming counterparts, to determine if there exist significant differences in the fate of ionizing radiation in these galaxies.  We find that the galaxy-to-galaxy variations in the SEDs is much larger than any systematic differences between starbursts and non-starbursts.
For example, we find no significant differences in the total absorption of ionizing radiation by dust, traced by the 24\mum, 70\mum, and 160\mum\ MIPS bands of the $Spitzer\ Space\ Telescope$, although the dust in starburst galaxies appears to be hotter than that of non-starburst galaxies.  We also observe no excess ultraviolet flux in the $GALEX$ bands that could indicate a high escape fraction of ionizing photons in starburst galaxies.  The small \Halpha\ fractions of the diffuse, warm ionized medium in starburst galaxies are apparently due to temporarily boosted \Halpha\ luminosity within the star-forming regions themselves, with an independent, constant WIM luminosity.  This independence of the WIM and starburst luminosities contrasts with WIM behavior in non-starburst galaxies and underscores our poor understanding of radiation transfer in both ordinary and starburst galaxies.
\end{abstract}

\section{Introduction} \label{s:intro}

One of the most urgent needs in understanding the evolution of cosmic structure is to clarify the origin and properties of the background ionizing radiation field.  In particular, quasar absorption-line systems are now a fundamental constraint on the development of large-scale structure in the cosmic web.  The statistical properties of the Lyman $\alpha$ absorbers appear to be plausibly reproduced by numerical simulations of a $\Lambda$CDM universe evolving from primordial fluctuations such as those in the cosmic microwave background \citep[e.g.][]{b:dave99, b:bi97}.  However, uncertainties in the background ionizing radiation field are a great obstacle in exploiting the wealth of absorber data \citep[e.g.][]{b:lidz06, b:sokasian03, b:madau99, b:rauch98}.  It seems clear that at redshifts $z \lesssim 3$, the UV background is mainly due to QSOs \citep[e.g.][]{b:madau99}.  However, at higher redshifts, the reduced QSO number density \citep{b:fan01, b:richards06} implies that another source, presumably stellar UV radiation from galaxies \citep{b:bouwens07}, is the dominant contributor.  The discovery of the large population of Lyman-break galaxies (LBGs) at these redshifts \citep[e.g.][]{b:steidel96} supports this scenario.

There is some evidence from both local starbursts \citep[e.g.][]{b:hoopes07} and in LBGs \citep[e.g.][]{b:shapley06} that at least some starbursts have high escape fractions \fesc\ for ionizing radiation.  However, most studies find that the \fesc\ is only a few percent, both locally \citep[e.g.][]{b:heckman01, b:leitherer95, b:grimes07} and at high redshift \citep[e.g.][]{b:wyithe10, b:steidel01, b:shapley06}.  \citet{b:siana10} find that no more than 8\%\ of galaxies at $z \sim 1.3$ can have relative escape fractions greater than 0.50, and that the average ionizing emissivity appears to decrease, approaching $z \sim 0$.
The low observed escape fractions are puzzling in view of predictions for higher values of \fesc.  For example, \citet{b:clarke02} predicted that above a threshold star-formation rate, the interstellar medium (ISM) is shredded by the superwind mechanical feedback, thereby opening avenues for the escape of Lyman continuum radiation.  
Furthermore, \citet{b:oey07} found that the fraction of \Halpha\ luminosity \fwim\ contributed by the diffuse, warm ionized medium (WIM), is much lower in starburst galaxies, defined as those galaxies with high \Halpha\ surface brightnesses.  While \fwim\ generally has a robust, universal value $\sim 0.5$ in star-forming galaxies, it is systematically lower in starbursts, by factors of 2 -- 5; \citet{b:oey07} suggested that the lower \fwim\ could be evidence of ISM density-bounding in the starbursts, suggesting the escape of ionizing radiation.

In the present study, we examine the multiwavelength spectral energy distributions of star-forming galaxies to better understand the fate of the ionizing radiation in starbursts vs non-starburst galaxies.  Observations in three mid- and far-infrared $Spitzer$ wavelengths, the MIPS 24\mum, 70\mum, and 160\mum\ bands, clarify the role of dust heating, while \Halpha\ and $GALEX$ NUV and FUV bands yield direct estimates of the ionizing stellar population.

\section{Data} \label{s:data}

{ 
\begin{figure*}[!t]
\plotone{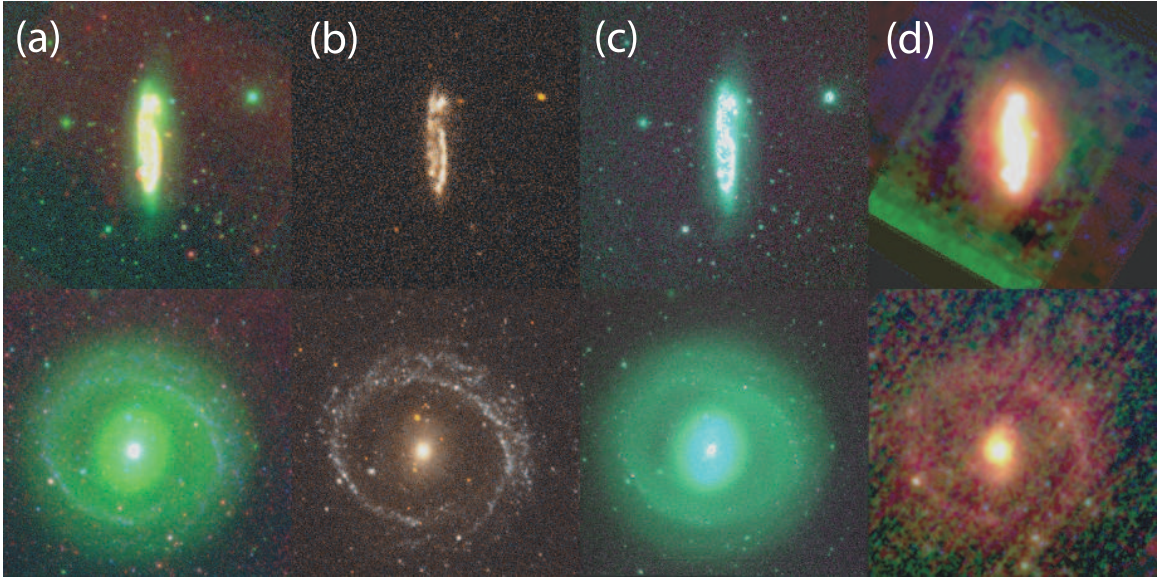}
\caption[Telescope-specific 3-color images]{3-color images, using NGC 1421 (top) and NGC 1291 (bottom) as examples of our starburst and non-starburst galaxies, respectively.  Columns shown are: (a) The primary composite image used in our masking: $Spitzer$ 24\mum\ flux is red, SINGG $R$ flux is green, and $GALEX$ NUV is blue; (b) the $GALEX$ bands: NUV flux is red, FUV flux is blue, and the average of the two bands is green; (c) the SINGG bands: R flux is blue, narrow-band flux is green, and the continuum-subtracted \Halpha\ flux is red; (d) the $Spitzer$ bands: 24\mum\ flux is blue, 70\mum\ flux is green, and 160\mum\ flux is red.  Images of this type for all 23 galaxies are available in the online edition of this paper.}
\label{f:4array}
\end{figure*}}
 
For our sample, we selected 23 local star-forming galaxies from the sample of 109 galaxies studied by \citet{b:oey07}, which correspond to the first data release of the Survey for Ionization in Neutral Gas Galaxies \citep[SINGG;][]{b:meurer06,b:hanish06}.  Since a strong differentiation was seen in \fwim\ between the starbursts and non-starbursts by \citet{b:oey07}, we chose 13 representative starburst galaxies and 10 non-starbursts from that work for further multi-wavelength study here.  We note that there are several definitions of starburst galaxies \citep[e.g.][]{b:lee09a}; we adopt the star formation intensity (SFI) criterion advocated by \citet{b:heckman05}.  Thus, \Halpha\ surface brightness is the corresponding observable quantity for the SFI, and starbrusts are defined as galaxies having \Halpha\ surface brightness within the \Halpha-derived half-light radius above a threshold $\log(\Sigma_{{\rm H}\alpha}\ [$erg\ s$^{-1}$\ kpc$^{-2}]) > 39.4$.  Similarly, the 10 non-starburst galaxies have $\Sigma_{{\rm H}\alpha}$ below this value.

We obtained new observations and archival data from the $Spitzer\ Space\ Telescope$.  These observations were taken with the MIPS camera, observing in the 24\mum , 70\mum , and 160\mum\ channels to estimate the thermal emission from hot and cold dust.  These MIPS bands are diffraction-limited, with effective resolutions of 6, 18, and 40 arcseconds, respectively.  The mosaic images have pixel scales of 2.45, 4.0, and 8.0 arcseconds per pixel, respectively, although this resolution is the result of a resampling algorithm applied to images with native pixel scales of up to 18 arcseconds per pixel for the 160\mum\ source images.  As a result, sources in the 70\mum\ and 160\mum\ images are significantly less resolved than those in our other bands.  This point spread function (PSF) results in a slight underestimation of total infrared fluxes due to aperture effects.  Aperture correction factors for the MIPS bands were estimated by \citet{b:dale09} to range from 1 -- 3\%, so this effect will not significantly alter our observed trends.

Of the 23 galaxies in our sample, we obtained new $Spitzer$ observational data for eight, as shown in Table~\ref{t:aps}.  Twelve additional galaxies had been fully observed by previous studies, four of which had archival data sets from multiple previous sources.  As a result, twenty of our sources (11 starbursts, 9 non-starburst galaxies) possessed the full array of MIPS observations, either from archival data or from our own requested observations.  The three galaxies with incomplete MIPS data were NGC 178 (lacked 24\mum\ and 160\mum\ data), NGC 1808 (lacked 160\mum\ data), and NGC 3365, which had no MIPS data at all due to the termination of the cold $Spitzer$ mission. 

{ 
\begin{deluxetable*}{l c c c c c c c c c c}
  \tablewidth{0pt}
  \tabletypesize{\small}
  \tablecaption{Aperture definitions, by galaxy \label{t:aps}}
  \tablehead{\colhead{Galaxy} &
             \colhead{RA (J2000)} &
             \colhead{Dec (J2000)} &
             \colhead{Distance} &            
             \colhead{$\log(\MHI)$} &
             \colhead{$v_{hel}$} &
             \colhead{$r_{50}$(\Halpha)} &
             \colhead{PA} &
             \colhead{$a/b$} &
             \colhead{$Spitzer$} &
             \colhead{$GALEX$} \\
             \colhead{} &
             \colhead{[h:m:s]} &
             \colhead{[d:m:s]} &
             \colhead{[Mpc]} &
             \colhead{[$\Msun$]} &
             \colhead{[km s$^{-1}$]} &
             \colhead{[arcsec]} &
             \colhead{[deg]} &
             \colhead{} &
             \colhead{reference$^{\dag}$} &
             \colhead{survey$^{\ddag}$}}
\startdata
\multicolumn{3}{l}{~~$Starburst\ Galaxies$} & & & & & & & & \\
ESO409-IG015 & ~0:05:31.7 & -28:05:49.2 & 10.24 & 8.27 & ~737$\pm$6 & 11.5 & 141.1 & 1.72 & (1) & (7) \\
NGC178 & ~0:39:08.2 & -14:10:26.4 & 19.8 & 9.40 & 1447$\pm$3 & 20.8 & ~~8.6 & 2.12 & (1) & (7) \\
NGC625 & ~1:35:03.1 & -41:26:13.2 & 4.45 & 8.09 & ~396$\pm$1 & 43.9 & ~94.0 & 2.96 & (2) & (7) \\
NGC922 & ~2:25:03.8 & -24:47:27.6 & 41.2 & 10.07 & 3082$\pm$5 & 30.7 & 176.0 & 1.09 & (1) & (7) \\
NGC1421 & ~3:42:29.3 & -13:29:20.4 & 27.7 & 9.85 & 2087$\pm$5 & 55.2 & 177.4 & 3.26 & (1) & (8) \\
NGC1487 & ~3:55:45.6 & -42:22:01.2 & 10.1 & 9.25 & ~848$\pm$1 & 31.3 & ~60.5 & 2.52 & (2) & (7) \\
NGC1510 & ~4:03:32.6 & -43:24:00.0 & 11.0 & 9.54 & ~~913$\pm$10 & ~3.3 & 129.4 & 1.13 & (2,3) & (9) \\
NGC1705 & ~4:54:14.2 & -53:21:39.6 & 6.46 & 7.96 & ~633$\pm$6 & 20.7 & ~44.1 & 1.38 & (3,4) & (9) \\
NGC1800 & ~5:06:25.2 & -31:57:18.0 & 9.7 & 8.54 & ~807$\pm$1 & 16.1 & 114.1 & 1.52 & (2) & (9) \\
NGC1808 & ~5:07:43.2 & -37:30:32.4 & 12.3 & 9.53 & ~995$\pm$4 & 39.0 & 134.3 & 2.10 & (5) & (7) \\
NGC5236 & 13:37:00.0 & -29:51:56.2 & 4.03 & 9.89 & ~513$\pm$2 & 122.0 & ~91.3 & 1.02 & (6) & (9) \\
NGC5253 & 13:39:55.4 & -31:38:27.6 & 2.54 & 8.22 & ~407$\pm$3 & 12.6 & ~44.8 & 2.01 & (2) & (9) \\
NGC7126 & 21:49:18.2 & -60:36:36.0 & 42.7 & 10.46 & 2981$\pm$2 & 17.8 & ~67.4 & 1.73 & (1) & (7) \\
\hline
\multicolumn{3}{l}{~~$Non$-$Starburst\ Galaxies$} & & & & & & & & \\
NGC1291 & ~3:17:18.0 & -41:06:32.4 & 10.1 & 9.41 & ~839$\pm$2 & 133.8 & ~76.7 & 1.17 & (3) & (9) \\
NGC1311 & ~3:20:06.7 & -52:11:06.0 & 5.95 & 8.25 & ~568$\pm$5 & 31.8 & ~38.6 & 2.60 & (2) & (10) \\
IC2000 & ~3:49:07.4 & -48:51:32.4 & 11.9 & 9.10 & ~981$\pm$4 & 54.5 & ~81.9 & 2.97 & (1) & (10) \\
NGC1512 & ~4:03:54.2 & -43:20:52.8 & 10.8 & 9.64 & ~898$\pm$3 & 69.6 & ~52.8 & 1.04 & (2,3) & (9) \\
NGC1744 & ~4:59:58.1 & -26:01:22.8 & 8.94 & 9.56 & ~741$\pm$2 & 89.0 & 164.6 & 1.68 & (2) & (11) \\
ESO486-G021 & ~5:03:19.7 & -25:25:26.4 & 10.3 & 8.60 & ~835$\pm$3 & 23.1 & ~97.0 & 1.64 & (2) & (8) \\
UGCA175 & ~9:43:36.5 & ~-5:54:43.2 & 30.8 & 9.36 & 2026$\pm$5 & 35.7 & ~20.4 & 1.29 & (1) & (7) \\
NGC3365 & 10:46:12.7 & +1:48:50.4 & 13.3 & 9.18 & ~986$\pm$1 & 45.8 & 157.2 & 3.25 & N/A & (7) \\
UGCA307 & 12:53:56.6 & -12:06:14.4 & 8.0 & 8.67 & ~821$\pm$1 & 21.7 & 142.6 & 2.14 & (1) & (7) \\
UGCA320 & 13:03:17.3 & -17:25:26.4 & 7.04 & 9.12 & ~742$\pm$2 & 42.0 & 117.4 & 3.36 & (2) & (7)
\enddata
\tablecomments{
\dag\ References: (1) Our requested data, $Spitzer$ GO-50332; (2) $Spitzer$ GO-40204 \citep{b:kennicutt07}; (3) $Spitzer$ GO-159 \citep{b:kennicutt03}; (4) $Spitzer$ GO-20528 (PI: Martin, C.); (5) $Spitzer$ GO-86 (PI: Werner, M.); (6) $Spitzer$ GO-59 (PI: Rieke, G.) \\
\ddag\ Surveys: (7) Guest Investigators Survey; (8) All-sky Imaging Survey; (9) Nearby Galaxy Survey; (10) Medium Imaging Survey; (11) LGAL Survey}
\end{deluxetable*}}

We used available \Halpha\ and $R$-band data from SINGG Release 1.  This survey uses an $R$ band continuum filter centered at 6508 \ang\, and a narrow-band \Halpha\ filter appropriate to the redshift of each galaxy to create a continuum-subtracted emission-line image.  The result is an accurate measure of \Halpha\ line flux, tracing the formation of high-mass stars within each galaxy.  SINGG images have a pixel resolution of 0.432 arcseconds per pixel, although as a ground-based survey, the images are limited by the effects of atmospheric seeing, which generally ranged from 1.0 to 1.5 arcseconds.  The SINGG field is approximately 14 arcminutes on each side; only one of our sources, NGC 5236 (Messier 83), was found to contain features extending beyond the bounds of its SINGG images.

We acquired archival $Galaxy\ Evolution\ Explorer$ ($GALEX$) data for all 23 galaxies.  These data consisted of two UV bands, 1516 \ang\ FUV (covering a range from 1350 \ang\ to 1750 \ang) and 2271 \ang\ NUV (ranging from 1750 \ang\ to 2800 \ang); while actual ionizing radiation remains shortward of both bands, these serve as approximate indicators of ionizing flux.  The detector used for both bands has a pixel scale of 1.5 arcseconds per pixel and a PSF FWHM of 4.5 -- 6 arcseconds, resulting in less detail than the SINGG images, but significantly more than the $Spitzer$ bands.  $GALEX$ images cover a far larger field than any of our other bands, a circular field with a diameter of approximately 1.2 degrees; as a result, the $GALEX$ NUV image served as the positional reference for the alignment algorithms used in our masking scripts.

$GALEX$ and SINGG data required no additional data reduction.  However, the $Spitzer$ observations required reprocessing with different parameters to ensure a better fit of the sky background for extended sources.  The method used to generate the intermediate Basic Calibrated Data (BCD) images, known as the Germanium Reprocessing Tools (GeRT), defaults to rejecting a 5$\times$5-pixel box around the center point of each object before calculating a sky profile.  We instead used Source Extractor \citep[v.2.5.0, ][]{b:bertin96} to identify the visibly extended features of each galaxy; this allowed us to then mask a larger area around each of these regions with a circular mask whose size is based on the semiminor axial length of each galaxy, resulting in a significantly improved model fit to the sky background.  Once this was done, the MOPEX software package was used to combine the modified BCD images into a single mosaic for each galaxy; only minor alterations to the default MOPEX process were needed, such as compensating for differing sky background levels in those few cases where we combined MIPS data from different archival sources.

All data processing beyond this stage was performed with IDL, the Interactive Data Language; all scripts were either custom-made or drawn from the IDL Astronomy User's Library ($\sf http://idlastro.gsfc.nasa.gov/$), more commonly known as ``astro-lib''.  For each galaxy, the various bands were aligned using IDL scripts contained in astro-lib.  Foreground objects were removed with a common mask; an initial mask was derived by applying Source Extractor to the $Spitzer$ 24\mum\ and $GALEX$ NUV images, and then iteratively modified by user input based primarily on a 3-color image similar to that of Figure~\ref{f:4array}(a), combining images from all three telescopes, while a series of alternate 3-color images (such as those telescope-specific images shown in Figure~\ref{f:4array}(b) through (d)) were used to evaluate any questionable objects.  The resulting mask was then scaled and rotated appropriately to match each input image's coordinates.  The $Spitzer$ 70\mum\ and 160\mum\ images were not significantly contaminated by foreground objects, as stellar continua are extremely faint at far-infrared wavelengths, so we did not use masks in these bands.  The end result was eight bands of data, each containing a pixel mask for foreground objects derived from the same source, allowing for a consistent flux measurement process.

The flux within each band was measured within a single elliptical aperture for each galaxy, whose center position, position angle, and axial ratio were taken directly from the SINGG project.  The fluxes were measured using the original input images, instead of the rotated and scaled versions used to generate the pixel masks.  However, the SINGG apertures were derived based on $R$ band flux and extended \Halpha\ emission, and so they tended to be large relative to the measurable size of the $Spitzer$ band emissions.  To maintain consistency, we limited our measurement of flux ratios to that part of each galaxy within the half-light radius measured in \Halpha, $r_{50}$, and this extended to derived quantities such as \Halpha\ equivalent width.

Corrections of internal extinction were drawn from relationships involving total infrared flux to \Halpha\ flux or UV flux, as appropriate; the $GALEX$ extinction prescriptions, Equations~\ref{e:ANUV} and \ref{e:AFUV}, were drawn from \citet{b:buat05}.  The extinction relation for \Halpha\ was from \citet{b:kennicutt09}, shown in Equation~\ref{e:AHa}, while the $R$-band extinction was assumed to be directly proportional to the \Halpha\ extinction by \citet{b:calzetti01}.  The extinction relations, in magnitudes, were
{\small
\begin{eqnarray}
x &=& \log(F_{\rm IR}/(\nu f_{\nu,{\rm NUV}})) \nonumber\\
A_{\rm NUV} &=& -0.0495x^{3} + 0.4718x^{2} + 0.8998x + 0.2269 \label{e:ANUV}\\
y &=& \log(F_{\rm IR}/(\nu f_{\nu,{\rm FUV}})) \nonumber\\
A_{\rm FUV} &=& -0.0333y^{3} + 0.3522y^{2} + 1.1960y + 0.4967 \label{e:AFUV}\\
A_{{\rm H}\alpha} &=& 2.5 \log(1.0 + 0.0024 F_{\rm IR}/(F_{{\rm H}\alpha})) \label{e:AHa}\\
A_{\rm R} &=& A_{{\rm H}\alpha} / 2.3 \label{e:AR}
\end{eqnarray}
}

As our infrared fluxes primarily represent thermal emission from various temperatures of gas and dust, no extinction corrections were applied to our three MIPS bands.  All of the above equations link dust extinction in each band to the total infrared flux.  To estimate the total $Spitzer$-derived dust emission, we calculated an integrated infrared flux from the model fits of \citet{b:dale07}, using their derived relationship
{\small
\begin{equation}
F_{\rm IR} = 1.559\ \nu f_{\nu,24\mu{\rm m}} + 0.7686\ \nu f_{\nu,70\mu{\rm m}} + 1.347\ \nu f_{\nu,160\mu{\rm m}}
\label{e:tir}
\end{equation}}
This expression, representing the total infrared flux between 3\mum\ and 1100\mum , corresponds primarily to the thermal emission of dust within each galaxy.  All fluxes used to determine extinction corrections were measured using the full SINGG-defined apertures instead of the reduced \Halpha-derived apertures used for all other fluxes in this paper, to maintain consistency with the studies which derived these relationships.

The extinction-corrected fluxes are given in Table~\ref{t:flux}; column 1 lists the galaxy ID, and columns 2 -- 8 give our measured fluxes in the seven bands we studied.  The \Halpha\ flux densities are calculated from the emission line fluxes by assuming a constant line width of 1.026 \ang, corresponding to the thermal broadening of a source at a temperature of 12,000 K.  While fluxes were directly available for several of these sources from the projects which provided many of the data images, we measured all fluxes ourselves to ensure consistent masks, apertures, and integration techniques were used.  The listed uncertainties are those due to sky measurement.  While many other sources of error exist, such as those of the [\NII] correction and continuum scaling ratio used in the derivation of \Halpha\ line flux, the total uncertainties are generally dominated by the extinction model used \citep{b:hanish06}.  As Equations~\ref{e:ANUV} -- \ref{e:AR} have no quoted uncertainties, we cannot easily quantify this effect.

{ 
\begin{deluxetable*}{l c c c c c c c}
  \tablewidth{0pt}
  \tabletypesize{\small}
  \tablecaption{Flux densities (log($f_{\nu}$ [Jy])$^{\dag}$) within $r_{50}(\Halpha)$, by galaxy \label{t:flux}}
  \tablehead{\colhead{Galaxy} &
             \colhead{FUV} &
             \colhead{NUV} &
             \colhead{$R$} &
             \colhead{\Halpha} &
             \colhead{24\mum} &
             \colhead{70\mum} &
             \colhead{160\mum}}
\startdata
\multicolumn{3}{l}{~~$Starburst\ Galaxies$} & & & & \\
ESO409-IG015 & -3.008$\pm$0.004 & -3.076$\pm$0.004 & -2.783$\pm$0.003 & -0.489$\pm$0.001 & -1.923$\pm$0.007 & -1.191$\pm$0.012 & -1.603$\pm$0.017 \\
NGC178 & -2.456$\pm$0.001 & -2.413$\pm$0.001 & -1.780$\pm$0.001 & -0.214$\pm$0.001 & N/A & -0.194$\pm$0.006 & N/A \\
NGC625 & -1.849$\pm$0.001 & -1.881$\pm$0.001 & -1.230$\pm$0.002 & ~0.581$\pm$0.001 & -0.526$\pm$0.004 & ~0.290$\pm$0.009 & ~0.202$\pm$0.011 \\
NGC922 & -1.822$\pm$0.001 & -1.819$\pm$0.003 & -1.364$\pm$0.003 & ~0.398$\pm$0.001 & -0.543$\pm$0.003 & ~0.530$\pm$0.005 & ~0.417$\pm$0.005 \\
NGC1421 & -1.831$\pm$0.002 & -1.780$\pm$0.003 & -1.032$\pm$0.001 & ~0.605$\pm$0.003 & -0.246$\pm$0.002 & ~0.810$\pm$0.005 & ~0.854$\pm$0.004 \\
NGC1487 & -1.953$\pm$0.001 & -1.959$\pm$0.001 & -1.440$\pm$0.002 & ~0.329$\pm$0.001 & -0.856$\pm$0.004 & ~0.164$\pm$0.006 & ~0.044$\pm$0.010 \\
NGC1510 & -2.758$\pm$0.001 & -2.862$\pm$0.001 & -2.169$\pm$0.001 & -0.231$\pm$0.001 & -1.409$\pm$0.001 & -0.980$\pm$0.006 & -1.323$\pm$0.015 \\
NGC1705 & -1.849$\pm$0.001 & -1.885$\pm$0.001 & -1.496$\pm$0.001 & ~0.275$\pm$0.001 & -1.467$\pm$0.004 & -0.316$\pm$0.003 & -0.599$\pm$0.005 \\
NGC1800 & -2.333$\pm$0.001 & -2.357$\pm$0.001 & -1.696$\pm$0.002 & -0.190$\pm$0.002 & -1.638$\pm$0.010 & -0.558$\pm$0.006 & -0.621$\pm$0.011 \\
NGC1808 & -2.078$\pm$0.001 & -1.832$\pm$0.001 & -0.615$\pm$0.001 & ~1.191$\pm$0.001 & ~0.703$\pm$0.001 & ~1.474$\pm$0.006 & N/A \\
NGC5236 & -0.389$\pm$0.001 & -0.385$\pm$0.001 & ~0.425$\pm$0.006 & ~2.020$\pm$0.010 & ~1.431$\pm$0.001 & ~2.144$\pm$0.006 & ~2.436$\pm$0.015 \\
NGC5253 & -1.378$\pm$0.001 & -1.594$\pm$0.001 & -1.311$\pm$0.003 & ~1.241$\pm$0.001 & ~0.561$\pm$0.001 & ~0.512$\pm$0.004 & ~0.158$\pm$0.005 \\
NGC7126 & -2.561$\pm$0.001 & -2.503$\pm$0.002 & -1.433$\pm$0.001 & ~0.084$\pm$0.001 & -0.907$\pm$0.002 & ~0.079$\pm$0.004 & ~0.024$\pm$0.004 \\
\hline
\multicolumn{3}{l}{~~$Non$-$Starburst\ Galaxies$} & & & & \\
NGC1291 & -2.091$\pm$0.004 & -1.844$\pm$0.003 & ~0.171$\pm$0.002 & ~0.607$\pm$0.031 & -0.575$\pm$0.083 & ~0.541$\pm$0.023 & ~0.904$\pm$0.018 \\
NGC1311 & -2.626$\pm$0.001 & -2.542$\pm$0.002 & -1.841$\pm$0.002 & -0.296$\pm$0.003 & -1.907$\pm$0.044 & -0.650$\pm$0.013 & -0.735$\pm$0.019 \\
IC2000 & -2.422$\pm$0.004 & -2.389$\pm$0.006 & -1.424$\pm$0.002 & -0.146$\pm$0.012 & -1.218$\pm$0.040 & -0.226$\pm$0.011 & ~0.129$\pm$0.005 \\
NGC1512 & -1.926$\pm$0.002 & -1.878$\pm$0.003 & -0.525$\pm$0.002 & ~0.417$\pm$0.020 & -0.528$\pm$0.033 & ~0.597$\pm$0.014 & ~0.954$\pm$0.017 \\
NGC1744 & -2.030$\pm$0.001 & -1.990$\pm$0.002 & -1.069$\pm$0.005 & ~0.265$\pm$0.084 & -1.098$\pm$0.090 & ~0.102$\pm$0.023 & ~0.530$\pm$0.018 \\
ESO486-G021 & -2.855$\pm$0.007 & -2.796$\pm$0.014 & -2.070$\pm$0.007 & -0.584$\pm$0.011 & -2.109$\pm$0.072 & -0.896$\pm$0.018 & -0.986$\pm$0.037 \\
UGCA175 & -2.617$\pm$0.006 & -2.579$\pm$0.011 & -1.697$\pm$0.002 & -0.422$\pm$0.017 & -1.480$\pm$0.032 & -0.431$\pm$0.023 & -0.193$\pm$0.008 \\
NGC3365 & -3.150$\pm$0.006 & -2.958$\pm$0.010 & -1.753$\pm$0.002 & -0.470$\pm$0.007 & N/A & N/A & N/A \\
UGCA307 & -3.071$\pm$0.002 & -3.078$\pm$0.004 & -2.604$\pm$0.007 & -0.579$\pm$0.003 & -2.261$\pm$0.047 & -1.507$\pm$0.066 & -1.435$\pm$0.033 \\
UGCA320 & -2.671$\pm$0.001 & -2.602$\pm$0.002 & -2.040$\pm$0.004 & -0.111$\pm$0.004 & -1.899$\pm$0.078 & -0.710$\pm$0.026 & -0.892$\pm$0.057
\enddata
\tablecomments{\dag\ All flux densities have been corrected for internal extinction, and quoted uncertainties only include sky measurement errors.}
\end{deluxetable*}}

\section{Comparison of Starburst and Non-starburst Galaxies} \label{s:results}

We calculate the star formation rates for each galaxy by three methods.  Our UV-derived and optically-derived SFRs use the relationships of \citet{b:kennicutt98}, and our infrared SFR uses that of \citet{b:calzetti07}.  These relationships are, for a Salpeter IMF, 
{\small
\begin{eqnarray}
{\rm SFR}_{\rm UV}[\Msun\ {\rm yr}^{-1}] &=& ~1.4 \times 10^{-28} L_{\nu,{\rm FUV}}[{\rm erg\ s}^{-1} {\rm Hz}^{-1}] \label{e:sfuv} \\
{\rm SFR}_{{\rm H}\alpha}[\Msun\ {\rm yr}^{-1}] &=& ~7.9 \times 10^{-42} L_{{\rm H}\alpha}[{\rm erg\ s}^{-1}] \label{e:sfha} \\
{\rm SFR}_{\rm IR}[\Msun\ {\rm yr}^{-1}] &=& 2.02 \times 10^{-38}\ [\nu L_{\nu,24\mu{\rm m}}[{\rm erg\ s}^{-1}]]^{0.885} \label{e:sfir}
\end{eqnarray}}

{ 
\begin{deluxetable*}{l c c c c c}
  \tablewidth{0pt}
  \tabletypesize{\small}
  \tablecaption{Derived flux values, by galaxy \label{t:flux2}}
  \tablehead{\colhead{Galaxy} &
             \colhead{\fwim$^{(1)}$} &
             \colhead{log($L_{\rm TIR}$[erg s$^{-1}$])} &
             \colhead{log(EW$_{{\rm H}\alpha}$[\ang])} &
             \colhead{log($\frac{{\rm SFR}(\Halpha)}{{\rm SFR}({\rm UV})}$)} &
             \colhead{$\frac{r_{50}}{r_{90}}$(\Halpha)}}
\startdata
\multicolumn{3}{l}{~~$Starburst\ Galaxies$} & & & \\
ESO409-IG015 & 0.15 & 40.80 & 2.29 & ~0.13 & 0.67 \\
NGC178 & 0.37 & N/A & 1.57 & -0.15 & 0.41 \\
NGC625 & 0.21 & 41.59 & 1.81 & ~0.04 & 0.57 \\
NGC922 & 0.29 & 43.68 & 1.76 & -0.17 & 0.77 \\
NGC1421 & 0.47 & 43.67 & 1.64 & ~0.04 & 0.70 \\
NGC1487 & 0.40 & 42.10 & 1.77 & -0.11 & 0.35 \\
NGC1510 & 0.22 & 41.25 & 1.94 & ~0.13 & 0.47 \\
NGC1705 & 0.52 & 41.16 & 1.77 & -0.27 & 0.35 \\
NGC1800 & 0.50 & 41.35 & 1.51 & -0.25 & 0.33 \\
NGC1808 & 0.19 & N/A & 1.81 & ~0.87 & 0.34 \\
NGC5236 & 0.31 & 43.52 & 1.59 & ~0.02 & 0.58 \\
NGC5253 & 0.11 & 42.22 & 2.55 & ~0.23 & 0.30 \\
NGC7126 & 0.54 & 43.30 & 1.52 & ~0.25 & 0.24 \\
~~$Mean^{2}$ & 0.33$\pm$0.15 & 42.2$\pm$1.1 & 1.81$\pm$0.31 & 0.06$\pm$0.30 & 0.47$\pm$0.17 \\
\hline
\multicolumn{3}{l}{~~$Non$-$Starburst\ Galaxies$} & & & \\
NGC1291 & 0.91 & 42.66 & 0.44 & ~0.30 & 0.42 \\
NGC1311 & 0.65 & 40.79 & 1.55 & -0.06 & 0.35 \\
IC2000 & 0.80 & 42.05 & 1.28 & -0.12 & 0.50 \\
NGC1512 & 0.72 & 42.77 & 0.94 & -0.05 & 0.42 \\
NGC1744 & 0.61 & 42.14 & 1.33 & -0.10 & 0.40 \\
ESO486-G021 & 0.70 & 41.02 & 1.49 & -0.12 & 0.56 \\
UGCA175 & 0.66 & 42.60 & 1.28 & -0.20 & 0.47 \\
NGC3365 & 0.71 & N/A & 1.28 & ~0.29 & 0.38 \\
UGCA307 & 0.59 & 40.37 & 2.03 & ~0.10 & 0.20 \\
UGCA320 & 0.57 & 40.86 & 1.93 & ~0.17 & 0.24 \\
~~$Mean^{2}$ & 0.69$\pm$0.10 & 41.7$\pm$0.9 & 1.35$\pm$0.46 & 0.02$\pm$0.18 & 0.39$\pm$0.11
\enddata
\tablecomments{(1) From \citet{b:oey07}; (2) Mean and standard deviation of each subsample.}
\end{deluxetable*}}

Quantities derived from our directly measured values are given in Table~\ref{t:flux2}.  Column 1 lists the galaxy ID, and column 2 shows the fraction of \Halpha\ luminosity \fwim\ observed in the diffuse, warm ionized medium, taken directly from \citet{b:oey07}.  Column 3 gives the total infrared luminosity for each galaxy based on Equation~\ref{e:tir}, column 4 gives the equivalent width of the \Halpha\ emission line within the \Halpha\ effective radius, column 5 shows the ratio of the UV-derived SFR from Equation~\ref{e:sfuv} to the \Halpha-derived SFR from Equation~\ref{e:sfha}, and column 6 gives the concentration index of our galaxies in \Halpha, defined as the ratio of the semimajor axis of the aperture containing 50\% of the \Halpha\ flux to that of the aperture containing 90\%.  Appropriate extinction corrections have been applied to the values of equivalent width (as this is just $F_{{\rm H}\alpha}/f_{\nu,R}$) and $\log({\rm SFR}(\Halpha)/\rm{SFR}(UV))$ given in columns 4 and 5.  The mean and standard deviation of each subsample is given for each variable, illustrating the differences between starburst and ordinary galaxies in quantities like \fwim\ and \Halpha\ equivalent width.

All of the extinction corrections in Equations~\ref{e:ANUV} -- \ref{e:AR} involve the ratio of $F_{\rm IR}$ to the flux at the appropriate wavelengths; while the IR luminosities given in Table~\ref{t:flux2} tend to be somewhat higher for starburst galaxies than for non-starbursts, the fluxes at other wavelengths increase by comparable amounts, resulting in similar distributions of extinction corrections.  For instance, the \Halpha\ extinction corrections of Equation~\ref{e:AHa} have a mean of $0.49\pm0.32$ magnitudes for starburst galaxies and $0.38\pm0.23$ magnitudes for non-starbursts, although this is the result of very inhomogenous populations; for instance, four out of thirteen starburst galaxies have \Halpha\ corrections above 0.95 magnitudes, but the remaining nine all possess corrections of 0.45 magnitudes or less.  The \Halpha\ corrections for ordinary star-forming galaxies range from 0.09 -- 0.71 magnitudes.  The $R$ band extinctions are, as given in Equation~\ref{e:AR}, directly proportional to the \Halpha\ extinctions and so follow a similar pattern.  The ultraviolet extinctions tend to be slightly higher than those of \Halpha, especially for starburst galaxies; in magnitudes, the NUV extinctions were $0.73\pm0.53$ for starbursts and $0.41\pm0.40$ for ordinary galaxies, while the FUV extinctions were $1.01\pm0.67$ for starbursts and $0.68\pm0.58$ for non-starbursts.


\subsection{Spectral Energy Distributions} \label{s:SED}

{ 
\begin{figure*}[!t]
\plotone{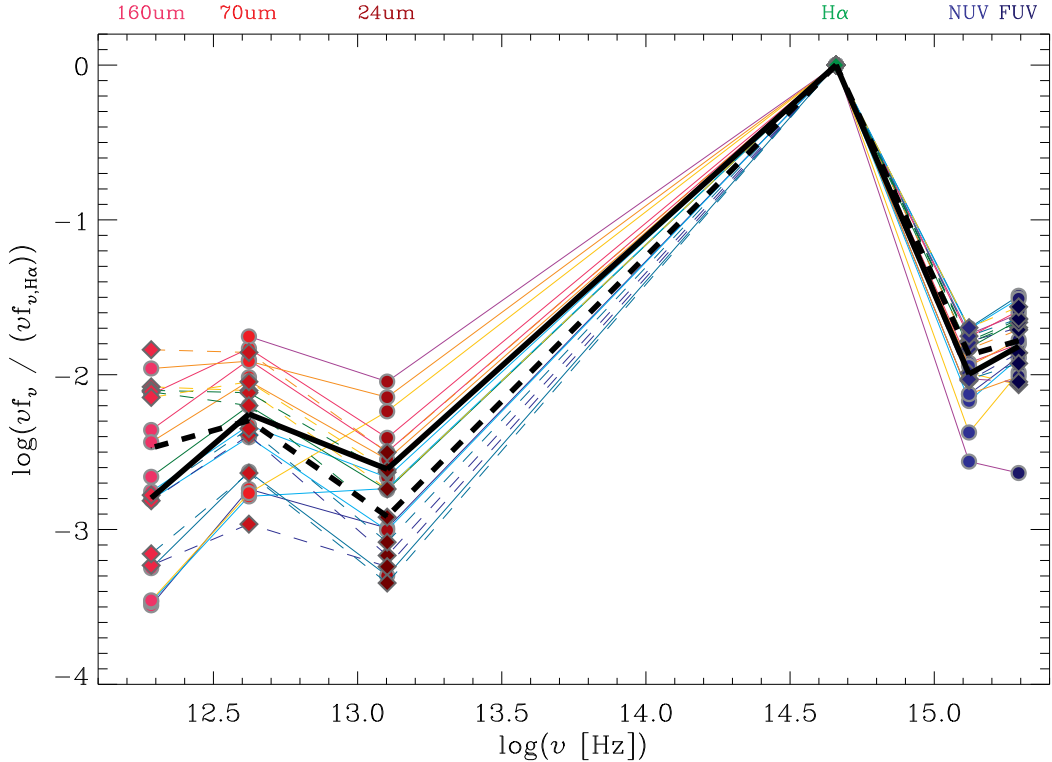}
\caption[Spectral energy distributions]{The SEDs for the galaxies in this sample.  Starburst galaxies use solid lines, while ordinary galaxies are represented by dashed lines, with line colors reflecting each galaxy's total IR flux ranging from high (red) to low (blue).  The heavier black lines are the averages for these two types.
}
\label{f:SED}
\end{figure*}}

Figure~\ref{f:SED} shows the spectral energy distribution (SEDs) for the galaxies in our sample, normalized by the \Halpha\ line flux, an indicator of star formation rate.  Starburst galaxies are shown with solid lines, and non-starbursts with dashed.  The mean composite SEDs for the starbursts and non-starbursts are shown with the heavier black solid and dashed lines, respectively, with error bars displaying the standard deviations of mean at each wavelength.  The variation between the individual galaxies is much larger than any systematic variation between the starburst and ordinary galaxy subsets.  That said, the starburst galaxies produced an average of 0.07 dex less ultraviolet flux per unit of \Halpha\ than their ordinary counterparts.  Likewise, for the 20 galaxies with complete MIPS data, the starburst galaxies produced an average of 0.03 dex less total infrared flux per unit \Halpha.  While at 24\mum\ we measured a difference of almost 0.3 dex between starbursts and ordinary galaxies, this gap was effectively negligible at 70\mum, and at 160\mum\ there was an anticorrelation of over 0.3 dex, resulting in a net discrepancy of 0.03 dex in total infrared flux from Equation~\ref{e:tir}, with starbursts again producing less flux per unit of \Halpha\ than their ordinary counterparts.  This difference is far less than the measurement uncertainties in each point, and is negligible compared to the statistical variation between the data points.  In both IR and UV, the individual galaxies' flux ratios often varied by up to 0.7 dex from the mean.  We note, however, that some of this scatter is introduced by galaxies with low \Halpha\ fluxes, which tend to have flux ratios further from the mean and significantly higher fractional uncertainties from sky subtraction.  Within our sample, these galaxies tended to have infrared-to-\Halpha\ ratios well below those of their larger counterparts, implying that our mean flux ratios might be slight underestimates.

{ 
\begin{figure}[!t]
\plotone{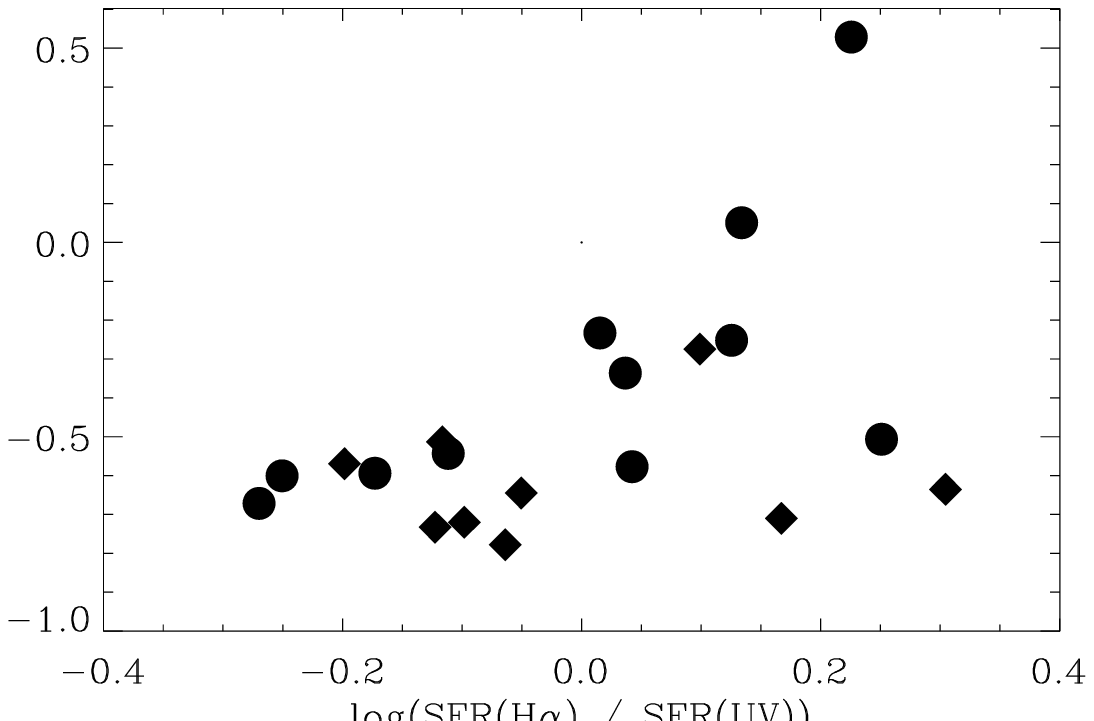}
\caption[$F_{24} / F_{70}$ versus SFR(\Halpha)/SFR(UV)]{Relationship between the ratio of 24\mum\ flux density to that at 70\mum\ and the ratio of \Halpha\ star formation rate to that measured using UV flux.  Circles denote starbursts, diamonds are ordinary galaxies.}
\label{f:2470}
\end{figure}}

Figure~\ref{f:2470} shows a modest correlation between the 24\mum/70\mum\ flux ratio and the \Halpha/UV star formation rate ratio.  Instead of showing the \Halpha/UV flux ratio itself, the abscissa in Figure~\ref{f:2470} shows the ratio of star-formation rates derived using Equations~\ref{e:sfuv} and \ref{e:sfha}, which are proportional to the SINGG \Halpha\ and $GALEX$ FUV fluxes respectively.  As noted above, Figure~\ref{f:SED} does show that the starbursts tend to have larger 24\mum/70\mum\ flux ratios; as this ratio should be expected to increase with temperature for purely thermal emission, this supports the observation that starbursts possess hotter dust temperatures \citep{b:draine07}, although this discrepancy would be far less without the presence of NGC 5253, the starburst galaxy with by far the highest 24\mum/70\mum\ ratio in our sample.  A correlation with the \Halpha/UV ratio links the 24\mum/70\mum\ ratio directly to the starburst activity itself.

The \Halpha/UV ratio can, in theory, be used to check whether ionizing photons are preferentially leaking into the IGM.  In practice, however, such an analysis is made difficult by the fact that many other processes can lead to an \Halpha/UV ratio that is lower than expected.  A number of studies based upon recent GALEX data \citep[e.g.][]{b:meurer09,b:lee09b,b:boselli09} as well as on earlier UV datasets \citep{b:bell01,b:sullivan04} have examined this issue in detail, and considered effects ranging from variations in the short-timescale star formation histories and dust attenuation, to the escape of ionizing photons and an IMF which is deficient in the most massive stars.  It is likely that most of the scatter of the \Halpha/UV ratio in Figure~\ref{f:2470} is due to non-constant star formation rates.  Nevertheless, it is notable that the average SFR ratios for the starburst and non-starburst samples are similar ($<\log($SFR$(\Halpha)/$SFR(UV))$>$ = $0.06\pm0.30$ and $0.02\pm0.18$, respectively, as shown in Table~\ref{t:flux2}), with similar ranges of values.  This argues that processes which lead to variations in the \Halpha/UV ratio, including Lyman continuum photon escape \citep{b:lee09b}, do not seem to be systematically dominant in the starburst population relative to ordinary star-forming galaxies.


\subsection{Radial Profiles}

{ 
\begin{figure*}[!t]
\plotone{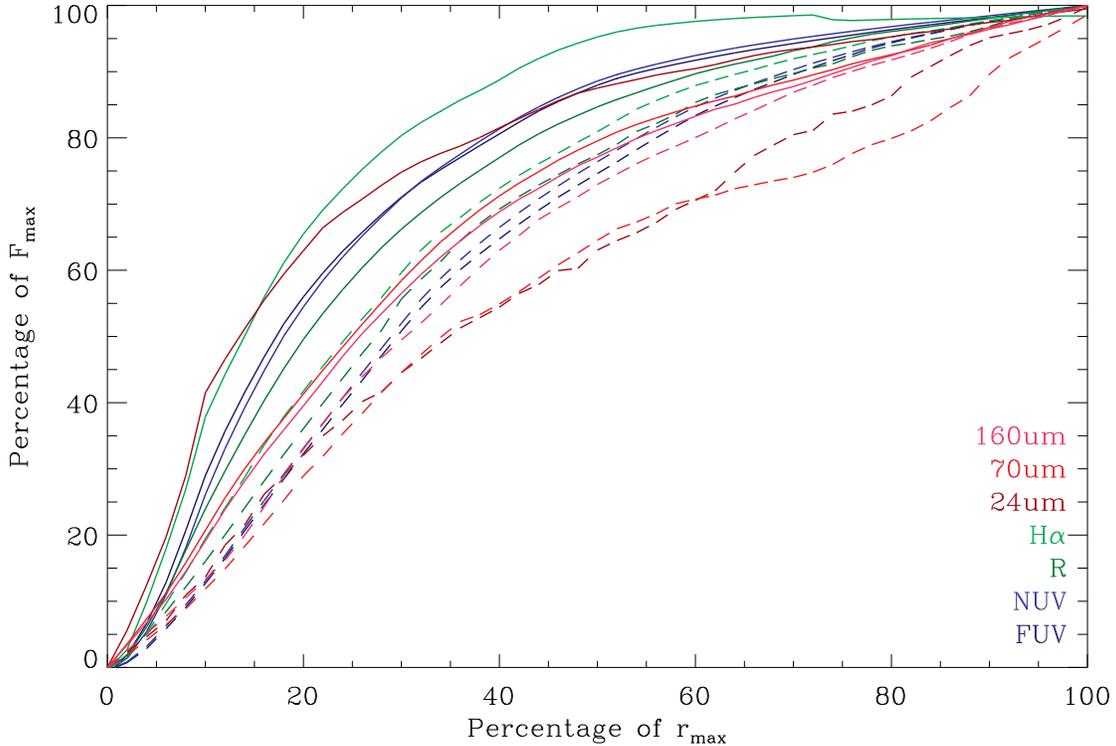}
\caption[Radial flux profiles]{Radial flux profiles in each wavelength band, using deconvolved images.  Solid lines are the means of the profiles for starburst galaxies, dashed lines represent ordinary star-forming galaxies.  The different bands are color-coded as shown.}
\label{f:rad}
\end{figure*}}

We also attempted to determine the mean radial distribution of the flux within each band, in order to compare the concentrations of star-forming regions to those of the hot and cold dust.  To do this, we deconvolved each IR and UV image by an appropriate PSF through a maximum entropy algorithm; this had the greatest effect on the $Spitzer$ MIPS images, due to their poor spatial resolutions.  We then measured the flux distribution of the resulting deconvolved image using the same annular method as in the earlier measurement of the total flux.  The results were plotted in Figure~\ref{f:rad} as the percentage of the total flux $F_{\rm max}$ as a function of the percentage of galaxy radius $r_{max}$, for each band.  The composite starburst and non-starburst radial profiles are shown by solid and dashed lines, respectively, with the values of $r_{max}$ determined from the SINGG \Halpha\ data for each galaxy.

The resulting profiles show several interesting trends.  The two ultraviolet bands are nearly identical in their radial distributions for both starbursts and ordinary galaxies, and both match well to the optical $R$-band distributions.  The link between \Halpha\ and 24\mum\ emission in starburst galaxies, suggested by Figure~\ref{f:2470}, is further confirmed by the coincidence of their radial distributions.  Figure~\ref{f:rad} shows that in starbursts, \Halpha\ and 24\mum\ emission are both strongly concentrated, with similar radial profiles, suggesting that the hot, young stellar population is directly responsible for the warmer dust traced by the 24\mum\ emission.  Similarly the $R$-band continuum correlates well to the UV bands for both starbursts and non-starbursts, with both acting as tracers of stellar population.

On the other hand, 24\mum\ flux for non-starburst galaxies is the least concentrated of any band.  Both 70\mum\ and 160\mum\ infrared fluxes were, on average, less concentrated than the other bands, but the data in these bands do not lend themselves well to measuring accurate profiles at small radii due to their low spatial resolutions, small fields of view, an imprecise deconvolution method, and uneven sky backgrounds.  While these factors are not significant enough to change the general trends seen in Figure~\ref{f:rad}, they do explain why the MIPS contours are more uneven than those of the SINGG and $GALEX$ bands.

{ 
\begin{figure}[!t]
\plotone{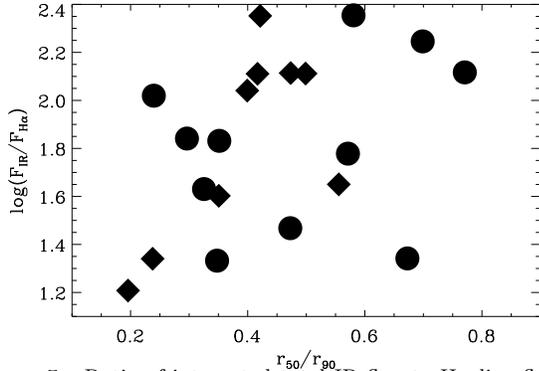}
\caption[Concentration]{Ratio of integrated total IR flux to \Halpha\ line flux from Table~\ref{t:flux}, as a function of concentration index $r_{50}/r_{90}$.  Circles denote starbursts, diamonds are ordinary galaxies.}
\label{f:concflux}
\end{figure}}

Given the observed correlation between the \Halpha\ and 24\mum\ radial profiles, we then evaluated whether the total IR flux implies significant absorption of ionizing radiation by dust in starburst galaxies.  While photons may be absorbed by dust at a wide range of wavelengths, the high-energy photons capable of ionizing hydrogen are more likely to be absorbed by even a moderate quantity of dust or gas and are unlikely to escape a galaxy entirely \citep{b:lee09b}.  If this hypothesis is correct, we would expect the total infrared flux to be larger in starburst galaxies relative to their \Halpha\ line flux, and the starbursts with more intense star formation would tend to have lower $r_{50}/r_{90}$ ratios, since their star formation activity is more concentrated.  Figure~\ref{f:concflux} shows the relationship between the IR/\Halpha\ flux ratio and the radial concentration of \Halpha.  We find no significant difference between starbursts and non-starbursts in this sample, either in the IR-to-\Halpha\ ratio or in concentration index; the mean $\log(F_{\rm IR}/F_{{\rm H}\alpha})$ for starbursts was $1.82\pm0.37$ while our non-starburst galaxies had $1.72\pm0.43$, and the distributions of $r_{50}/r_{90}$ values are given in Table~\ref{t:flux2}.

\subsection{The Warm, Ionized Medium}

{ 
\begin{figure}[!t]
\plotone{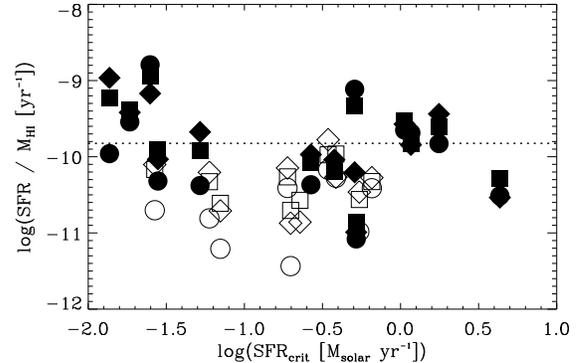}
\caption[Star formation rates by band]{Plotting SFR/\MHI\ versus the critical SFR needed to shred the ISM.  Circles represent IR, diamonds are UV, squares are \Halpha\, and the dotted line is the \HI-based \SFRcrit\ derived from equation~\ref{e:sfrcrit}.  Filled symbols represent starbursts, open symbols are ordinary galaxies.}
\label{f:sfrs}
\end{figure}}

As mentioned in \S~\ref{s:intro}, \citet{b:oey07} suggested that the small WIM fractions \fwim\ seen in \Halpha\ observations of starburst galaxies could be caused by the escape of ionizing radiation.  If the starburst activity yields enough ionizing radiation that all of the diffuse ISM is photoionized, then even if the starburst luminosity increases, the total WIM luminosity remains constant or decreases, causing \fwim\ to decrease.  High \fesc\ is furthermore plausible in starburst galaxies, based on theoretical predictions.  For example, \citet{b:clarke02} suggested the existence of a threshold star formation rate above which a galaxy's ISM is shredded by mechanical feedback, allowing UV radiation to escape freely:
\begin{equation}
\SFRcrit[\Msun\ {\rm yr}^{-1}] = 0.15 \frac{M_{{\rm ISM},10}\ \tilde{v}^{2}_{10}}{f_{d}}
\label{e:sfrcrit}
\end{equation}
where $M_{{\rm ISM},10}$ is the ISM mass in units of 10$^{10}$ \Msun, $\tilde{v}_{10}$ is the thermal velocity dispersion of the ISM in units of 10 km s$^{-1}$, and $f_{d}$ is a geometric correction factor for disk galaxies.  This relation results from a simple criterion that balances the supernova mechanical energy resulting from star formation against the total ISM thermal energy.  If the former dominates, the ISM is shredded, a galactic outflow is generated, and ionizing photons escape.  As detailed in \citet{b:oey07}, for this discussion the ISM mass can be approximated by the \HIPASS-derived \HI\ mass \MHI , the thermal velocity dispersion of the ISM is assumed to be a constant 10 km s$^{-1}$.  $f_{d}$ is assumed to be 0.1 for our entire sample to account for a flattened, disk geometry; as a result of this geometry the star formation intensity, shown to relate to galactic outflow by \citet{b:heckman02}, can be assumed to be directly proportional to the total star formation rate of each galaxy.  Taken together, these assumptions make \SFRcrit\ directly proportional to the \HI\ mass.

The SFR derived from each band, according to Equations~\ref{e:sfuv}, \ref{e:sfha}, and \ref{e:sfir}, are shown in Figure~\ref{f:sfrs}, normalized by \HI\ mass.  The starburst galaxies are the only ones with star formation rates consistently surpassing the \SFRcrit\ line given by Equation~\ref{e:sfrcrit}, with nearly all starbursts having at least one tracer of star formation fall within 0.2 dex of this critical threshold.  In contrast, almost all ordinary star-forming galaxies fall at least 0.3 dex below the \SFRcrit\ line, with only UGCA 175 having even a single tracer of star formation with a value above the critical threshold.  This confirms our earlier findings based only on \Halpha\ observations, that the starburst galaxies are expected to have high \fesc, although we caution that the crude relation in Equation~\ref{e:sfrcrit} needs refinement.  Another possibility is that \SFRcrit\ is underestimated if the mass of the ISM is predominantly molecular.

{ 
\begin{figure}[!t]
\plotone{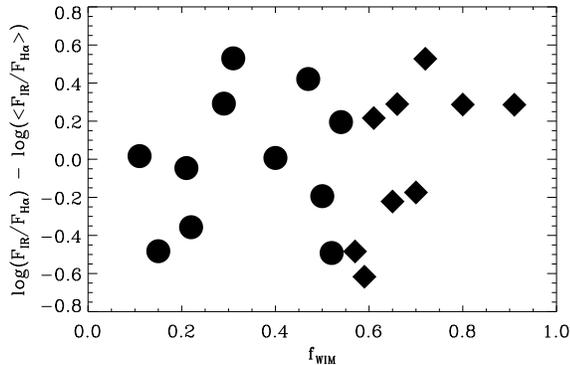}
\caption[Deviation in \fwim]{Plotting the deviation from the mean of the (IR/\Halpha) ratio for each galaxy in our sample against the fraction of warm, ionized medium derived from our earlier work.  Circles represent starbursts, diamonds are ordinary galaxies.}
\label{f:fwim}
\end{figure}}

However, in \S~\ref{s:SED}, we saw that the \Halpha-derived star-formation rates for starburst galaxies are not systematically suppressed relative to those determined from UV or IR data, as would be expected for high \fesc.  The SFR calibrations are based on total luminosities, rather than star-formation intensity, which is the differentiating factor here between starbursts and massive star-forming galaxies.  Thus, the similarity between starburst and non-starburst SEDs suggests no significant \fesc\ in the starbursts.

We therefore seek an alternative explanation for the low \fwim\ observed in the starburst galaxies of \cite{b:oey07}.  One possibility is a disproportionate absorption of Lyman continuum photons by dust in the starbursts \citep{b:lee09b}.  However, Figure~\ref{f:fwim} shows no correlation between the deviation from the mean in the IR/\Halpha\ flux ratio and the fraction of diffuse, warm ionized medium (\fwim), which we know to be lower in starburst galaxies.  This is also expected from our inference above, that any inferred difference between the starbursts and non-starbursts is insignificant for this sample.

{ 
\begin{figure}[!t]
\plotone{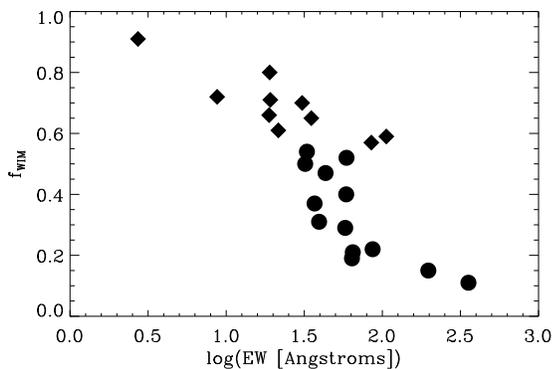}
\caption[EW versus \fwim]{\Halpha\ equivalent width as a function of the warm ionized medium fraction (\fwim).  Circles denote starbursts, diamonds are ordinary galaxies.}
\label{f:ewfwim}
\end{figure}}

A more plausible explanation for the high WIM fractions in starbursts is suggested by Figure~\ref{f:ewfwim}, which shows \fwim\ as a function of \Halpha\ equivalent width: a strong anticorrelation is apparent.  The \Halpha\ equivalent width is essentially a measure of the starburst age, since the line emission is strongest in the youngest regions, while the stellar continuum changes on much longer timescales and increases with age.  Thus, the anticorrelation in Figure~\ref{f:ewfwim} strongly suggests that starburst galaxies simply have a temporarily augmented, intense star formation, which increases the total \Halpha\ luminosity concentrated within the \HII\ regions.  In contrast, {\it the luminosity of the diffuse WIM remains largely independent of the starburst activity,} and so its fractional contribution to a starburst galaxy's total \Halpha\ luminosity decreases.

Table~\ref{t:flux2} shows that, as expected, the mean starburst has an \Halpha\ equivalent width 0.46 dex higher than that of the mean ordinary galaxy, corresponding to a factor of three increase in mean star formation intensity.  In fact, there was very little overlap between the two samples; none of our starburst galaxies had equivalent widths below $\log(EW(\Halpha)) < 1.51$, while only two out of ten ordinary galaxies had values greater than 1.55.

This apparent disconnect between the starburst and WIM luminosities contrasts strongly with what is observed in ordinary star-forming galaxies.  It is well-known that ordinarily, the WIM comprises about half of the total \Halpha\ luminosity in star-forming galaxies, regardless of galaxy luminosity, morphology, or star-formation intensity \citep[e.g.][]{b:walterbos98}.  \citet{b:oey07} studied a much larger sample of star-forming galaxies than had previously been examined, confirmed this result, but with the notable exception of the starburst galaxies.  The physical basis for the constant WIM fraction is not understood, and the variation from this behavior by starbursts may be an important clue.  More work is needed to understand the radiation transfer of ionizing photons in {\it both} starbursts and ordinary star-forming galaxies.

\section{Conclusions} \label{s:conc}

Our multiwavelength comparison of starburst and non-starburst galaxies shows that on average, the SEDs of the starburst galaxies, defined by star-formation intensity, do not differ significantly from those of their ordinary star-forming counterparts.  Within each band, individual starburst and non-starburst galaxies varied by up to 0.7 dex in their flux ratios, and only in 24\mum\ and 160\mum\ fluxes were the gaps between the mean flux ratios of the starburst and non-starburst galaxies larger than the statistical uncertainty in each.  As a result, our data do not support the hypotheses that ionizing radiation is either escaping or being absorbed by nearby dust in our sample of starbursts, in significantly larger fractions than would happen in ordinary star-forming galaxies.

We do find evidence that dust in the starburst galaxies is somewhat hotter, as shown by the higher 24/70\mum\ flux ratios and the extremely concentrated 24\mum\ radial profile of starbursts.  The coincidence of the \Halpha\ and 24\mum\ radial profiles in starburst galaxies strongly suggests that the young stellar population traced by \Halpha\ flux is responsible for the increase in hot dust traced by 24\mum\ infrared flux.  However, the overall distribution of the radiant energy emission in this \Halpha-selected sample shows the starbursts and non-starburst galaxies to be broadly similar; while starburst galaxies appear to have substantially higher amounts of hot dust traced through 24\mum\ flux, they have a correspondingly reduced amount of cold dust detected through 160\mum\ flux, with total infrared fluxes comparable to their ordinary star-forming counterparts.

We find no evidence that the low WIM fractions in starburst galaxies are caused by higher \fesc.  Furthermore, there is no correlation of infrared excess with \fwim, implying that dust absorption of the ionizing radiation also cannot explain this effect.  Instead, an inverse correlation of \Halpha\ equivalent width and \fwim\ (Figure~\ref{f:ewfwim}) strongly suggests that the WIM luminosity is independent of the starburst activity, and so the WIM contribution shrinks when the \Halpha\ luminosity is temporarily boosted by the starburst.  This behavior contrasts strongly with the relationship between star formation and WIM luminosity in ordinary star-forming galaxies.

In summary, this study is consistent with those that found low \fesc\ from starbursts.  Why these galaxies with extreme star formation have low \fesc\ is not understood, nor do we know why the WIM luminosity correlates with SFR in ordinary galaxies, even though it is now apparent these do not correlate in starburst galaxies.  These mysteries emphasize that our physical understanding of radiation transfer for ionizing photons is extremely limited.  We also caution that our sample is limited in parameter space, especially in terms of the magnitude of starburst activity, and additional studies, including those of more extreme systems, are needed to fully understand the circumstances that allow for high escape fractions of ionizing radiation.
\\
\acknowledgments

Support for this work was provided by NASA through contracts $Spitzer$ GO-50332 and NASA-ADP NNX08AJ42G.  We are grateful to Chad Engelbracht and the Spitzer Science Center for helpful discussions on the MIPS data reduction.  We also appreciate useful discussions with Daniela Calzetti, and comments from the anonymous referee.


\clearpage

\end{document}